\documentstyle[preprint,aps]{revtex}
\begin{document}
\draft
\title{Temperature induced transitions between insulator, metal, and
quantum Hall states}
\author{S.\ V.\ Kravchenko, Whitney Mason, and J.\ E.\ Furneaux}
\address{Laboratory for Electronic Properties of Materials and
Department of Physics and Astronomy, University of Oklahoma, Norman,
Oklahoma 73019}
\author{J.~M.~Caulfield and J.~Singleton}
\address{Clarendon Laboratory, Department of Physics, University of
Oxford, Parks Road, Oxford OX1~3PU, United Kingdom}
\author{V.~M.~Pudalov}
\address{Institute for High Pressure Physics, Troitsk, 142092 Moscow
district, Russia}
\maketitle
\begin{abstract}
We report the observation of temperature induced transitions between
insulator, metal, and quantum-Hall behaviors for transport
coefficients in the very dilute high mobility two-dimensional
electron system in silicon at a magnetic field corresponding to
Landau level filling factor $\nu=1$. Our data show that as the
temperature decreases, the extended states at $\nu=1$ (above the
Fermi level at higher temperature so that the system is insulating)
sink below the Fermi energy, so that the quantum Hall effect occurs.
As the extended states cross the Fermi level, the conductivity has a
temperature dependence characteristic of a metallic system.
\end{abstract}
\pacs{PACS 73.40.Hm, 73.40.Qv, 71.30.+h}
\narrowtext
In the study of two dimensional electron systems (2DES), there have
been considerable interest in transitions between insulating and
quantum Hall (QH) states. Thus far such transitions have only been
observed as the magnetic field and/or the electron density are
changed. However, in this paper we present experimental evidence that
transitions between insulator, metal, and QH-like behaviors for
transport coefficients can also be observed at fixed field and
carrier density as the {\em temperature} is decreased. In particular,
we have observed 2DES {\em delocalization} (insulator to metal
transition) with decreasing temperature, whereas increasing
localization is the usual consequence of lowering the temperature for
a 2DES.

It was reported recently both for high-mobility silicon inversion
layers \cite{diorio90} and for low-mobility GaAs/(Al,Ga)As
heterostructures \cite{jiang93,wang94} that 2DES, strongly localized
in zero magnetic field ($B=0$), can nevertheless manifest the integer
QH effect. References \cite{kravchenko91,diorio92,pudalov93} discuss
this magnetic-field-induced transition from a many-body point of
view, whereas Refs.\ \cite{jiang93,wang94} ignore the effects of
electron-electron interactions and consider the arguments of
Khmelnitskii \cite{khmelnitskii84} and Laughlin \cite{laughlin84}
along with the global phase diagram proposed by Kivelson, Lee, and
Zhang \cite{kivelson92} as the basis for the insulator-QH transition.
According to Refs.\ \cite{khmelnitskii84,laughlin84}, extended
states, which lie above the Fermi energy ($E_F$) at $B=0$ ({\em i.e.}
making the system insulating) decrease in energy as $B$ is increased
[see inset in Fig.~\ref{1}~(a)] and may sink below $E_F$
\cite{floating}. At zero temperature, the diagonal conductivity
($\sigma_{xx}$) is zero except when extended states are at $E_F$. As
long as $E_F$ lies below the lowest extended state, the Hall
conductivity ($\sigma_{xy}$) is also zero and the system is
insulating. For each band of extended states below $E_F$,
$\sigma_{xy}$ increases by $e^2/h$, providing the next integer QH
state. In high magnetic fields, the extended states approximately
follow half-filled Landau levels with filling factors
$\nu=n_s/(eB/ch)=1/2$, 3/2, 5/2 {\it etc} (here $n_s$ is the electron
density, $e$ is the electron charge, $c$ is the speed of light, and
$h$ is the Plank constant). If temperature is not zero, $\sigma_{xx}$
usually displays activated behavior except when there are extended
states at $E_F$ with the activation energy equal to the energy
difference between $E_F$ and the nearest extended state.

Magnetic-field-induced transitions between insulating and QH
groundstates have also been investigated in very dilute high-mobility
GaAs/(Al,Ga)As heterostructures in the extreme quantum limit (see,
{\em e.g.}, Ref.\ \cite{jiang91} and references therein) and in Si
inversion layers around $\nu=1$, 2, and 6 (Ref.\
\cite{diorio92,pudalov93}). In some papers
\cite{kravchenko91,diorio92,pudalov93,jiang90,goldman90,santos92} the
fact that the insulating state can be disrupted by integer or
fractional QH resistivity minima was discussed in terms of the
formation of a pinned electron solid melting at integer or fractional
$\nu$. However, the global phase diagram \cite{kivelson92} can also
explain magnetic-field-induced transitions between QH (at $\nu=1$,
1/3, or 1/5) and insulating groundstates without invoking collective
effects; though, recently observed direct transitions from $\nu=2/7$
and 2/5 (Refs.\ \cite{santos92,manoharan94}) and $\nu=6$ (Ref.\
\cite{pudalov93}) to insulating states are contradictions of the
global phase diagram. At very low $n_s$, the QH effect breaks down
even at integer or fractional $\nu$ \cite{diorio90,santos92};
nevertheless $\rho_{xx}$ still has minima at these filling factors.

Here we consider the $\nu=1$ integer QH effect at the border of it's
existence, at very low $n_s$. There is definitely no electron solid
for these conditions. We report experimental evidence for transitions
from insulating to metallic and QH types of behavior of transport
coefficients at constant filling factor $\nu=1$ {\em as the
temperature is varied}. We will define a system as insulating if its
conductivities exhibit insulator-like temperature behavior, {\em
i.e.}, $d\sigma_{xx}/dT$, $d\sigma_{xy}/dT>0$ with
$\sigma_{xy}<\sigma_{xx}<e^2/h$. Similarly we define ``metallic''
behavior as $d\sigma_{xx}/dT$, $d\sigma_{xy}/dT<0$.  Finally, a QH
state is characterized by $\sigma_{xy}\rightarrow e^2/h$ and
$\sigma_{xx}\rightarrow0$ as the temperature is lowered. The observed
transitions from insulating behavior can be explained by assuming
that at $\nu=1$, the lowest band of extended states, which lies above
the Fermi energy at high temperature as expected for an insulator,
drops as the temperature is lowered. The system behaves as a metal as
it passes through the Fermi level, and finally displays the QH effect
after the band of extended states sinks below $E_F$.

Two samples from wafers with different mobilities have been studied:
Si-14, which has a maximum mobility ($\mu_{max}$) of $1.9\times10^4$
cm$^2$/Vs, and Si-22, which has $\mu_{max}=3.5\times10^4$ cm$^2$/Vs.
The samples are rectangular with a source to drain length of 5~mm, a
width of 0.8~mm, and an intercontact distance of 1.25~mm. Resistances
were measured using a four-terminal low-frequency (typically 8 Hz) AC
technique using cold amplifiers with input resistances
$>10^{14}$~$\Omega$ installed inside the cryostat. The output of
these amplifiers was connected to a standard lock-in amplifier. Great
care was taken to ensure that all data discussed here were obtained
where the $I-V$ characteristics are linear.

The dependencies of $\rho_{xx}$ and $\rho_{xy}$ (both are in units of
$h/e^2$) on magnetic field for three different temperatures are shown
in Fig.~\ref{1}~(a). At the highest temperature, $\rho_{xx}(B)$ is
flat up to $B\approx4$~T and lies well above $h/e^2$. As the
temperature decreases, minima near integer filling factors $\nu=1$
and 2 and maxima at intermediate filling factors $\nu\sim1.5$ and 2.7
appear. The Hall resistance is almost $T$-independent; at low
temperature, narrow QH plateaux start to develop near $\nu=1$ and 2
\cite{hall}. Note that as the temperature is decreased from 1.82~K to
942~mK, $\rho_{xx}$ at $\nu=1$ decreases while remaining larger than
$\rho_{xy}=h/e^2$; this corresponds to $d\sigma_{xx}/dT<0$
characteristic of a {\em metallic} state and reflects {\em
delocalization} with decreasing temperature, whereas $\rho_{xx}$
increases at $B=0$, indicating an insulating groundstate. At still
lower temperature, $\rho_{xx}$ at $\nu=1$ sinks below
$\rho_{xy}=h/e^2$ penetrating into the QH region.

Before proceeding further, we must demonstrate that the observed
metallic-like decrease in $\rho_{xx}$ near $\nu=1$ is a fundamental
effect of the whole 2DES and not an artifact of special current paths
such as, {\em e.g.}, edge currents \cite{mceuen90}. Furthermore, in
principle the current distribution in a 2DES at low temperatures can
be inhomogeneous. If this is the case, it is difficult to draw
conclusions about the behavior of $\sigma_{xx}$ using $\rho_{xx}$
data. To check these points and to obtain information about
$\sigma_{xx}$ directly, we have  measured the impedance between the
2D channel and the metallic gate using an $RC$ bridge. The real part
of the bridge imbalance signal is proportional to inverse
$\sigma_{xx}$ averaged over the sample area. The magnetic field
dependence of the signal proportional to $\sigma_{xx}^{-1}$ is shown
in Fig.~\ref{1}~(b). One can see that it is qualitatively similar to
$\rho_{xx}(B)$ as expected for the case of $\rho_{xx}>\rho_{xy}$.
[For ``normal'', high-$n_s$ QH effect, where $\rho_{xx}<\rho_{xy}$,
the dependence is qualitatively different: at integer filling factors
$\nu=1$ and 2, $\sigma_{xx}^{-1}$ has extreme {\em maxima} instead of
minima, see Fig.~\ref{1}~(b) inset]. Again, as the temperature is
lowered, $\sigma_{xx}$ increases at $\nu=1$, showing metallic
behavior, and decreases at $B=0$, showing insulating behavior. The
similarity of the independently determined $\rho_{xx}(B)$ and
$\sigma_{xx}^{-1}(B)$ shows that the current distribution can be
considered homogeneous and that we can therefore calculate
$\sigma_{xx}$ and $\sigma_{xy}$ from the data for $\rho_{xx}$ and
$\rho_{xy}$.

Figure~\ref{2} shows the temperature dependence of $\rho_{xx}$ at
$\nu=1$ for four different electron densities. For the lowest $n_s$,
$\rho_{xx}(T)$ always lies above $h/e^2$ and monotonically increases
as the temperature is decreased showing an activated temperature
behavior (see inset). On the other hand, for the highest electron
density, $\rho_{xx}$ always lies below $h/e^2$ and monotonically
decreases as the temperature is decreased below $T\approx3$~K showing
QH type of behavior. But for the two middle $n_s$, $\rho_{xx}(T)$ are
nonmonotonic, increasing exponentially at high temperatures (see
inset) and decreasing at lower ones. This nonmonotonic behavior
implies, as shown below, temperature induced transitions from
insulator-like to metallic and QH types of behavior.

The temperature dependencies of $\sigma_{xx}$ and $\sigma_{xy}$,
computed from the data for $\rho_{xx}$ and $\rho_{xy}$, are shown in
Fig.~\ref{3}. Figure~\ref{3}~(a) shows temperature behavior of
$\sigma_{xx}$ and $\sigma_{xy}$ for the case of ``ordinary'',
``high-$n_s$'' QH effect. Diagonal conductivity, which is always less
than $e^2/2h$, {\em monotonically} decreases, and $\sigma_{xy}$,
which is always higher than $e^2/2h$, {\em monotonically} increases
as the temperature is decreased. Neither metallic nor insulator-like
behavior is observed at $n_s\gtrsim1.3\times10^{11}$~cm$^{-2}$ for
any temperature. However, at slightly lower electron density
[Fig.~\ref{3}~(b)], both components of conductivity are no longer
monotonic functions of~$T$:

(i) at $T\gtrsim2.5$~K (to the left of the first vertical dotted
line), $\sigma_{xy}<\sigma_{xx}<e^2/h$ and both diminish with
diminishing $T$ which is characteristic for an insulating state.
Furthermore, according to Khmelnitskii~\cite{khmelnitskii84},
$\sigma_{xy}$ in units of $e^2/h$ is a ``counter'' of the number of
bands of extended states below $E_F$, and because $\sigma_{xy}\ll
e^2/h$ at $T\gtrsim2.5$~K, there are no extended states below $E_F$
which confirms an insulating state in this temperature region.

(ii) at $1\lesssim T\lesssim2.5$~K (between the two vertical dotted
lines), both $\sigma_{xx}$ and $\sigma_{xy}$ increase with decreasing
$T$, indicating ``metallic'' behavior. Both also reach $e^2/2h$, the
value expected for an extended state
($\sigma_{xx}^0\sim\sigma_{xy}^0=e^2/2h$ at $\nu=1/2$ and $T=0$
\cite{ando82}).

(iii) below $T\approx1$~K (to the right of the second dotted line),
$\sigma_{xx}$ again tends to zero as $T\rightarrow0$ while
$\sigma_{xy}$ approaches $e^2/h$. The value of $\sigma_{xy}$ now
corresponds to one band of extended states below $E_F$; this is a QH
state.

At even lower $n_s$ [Fig.~\ref{3}~(c)], both $\sigma_{xx}(T)$ and
$\sigma_{xy}(T)$ change their slope from ``insulator-like'' to the
``metallic'' at $T\approx1$~K but $\sigma_{xy}$ always remains
$<e^2/2h$ and the QH conditions are never obtained for this $n_s$.
Finally, for the lowest $n_s$ [Fig.~\ref{3}~(d)], both $\sigma(T)$
decrease monotonically down to the lowest temperatures while
$\sigma_{xy}<\sigma_{xx}<e^2/h$; for this $n_s$, the system always
remains insulating. In summary, for
$n_s\gtrsim1.3\times10^{11}$~cm$^{-2}$ we have observed that the
system remains in the QH state regardless of temperature. For
$n_s=9.5\times10^{10}$~cm$^{-2}$, we have observed three different
types of behavior at $\nu=1$: insulator-like at higher temperatures,
metallic at intermediate, and QH at $T\rightarrow0$. For lower
$n_s=8.6\times10^{10}$~cm$^{-2}$, we have observed a transition from
insulating to metallic types of behavior without a further transition
to QH behavior. For even lower $n_s=8.1\times10^{10}$~cm$^{-2}$, the
system remains insulating.

We can understand this effect phenomenologically by examining the
position of the energy of the lowest extended state at $T=0$.
According to Refs.\ \cite{khmelnitskii84,laughlin84}, this is
\begin{equation}
E_c=\frac{1}{2}\,\hbar\Omega_c\left[1+(\Omega_c\tau)^{-2}\right],
\end{equation}
where $\Omega_c$ is the cyclotron frequency and $\tau$ is the
relaxation time. Therefore, at constant $B$ ($\Omega_c=$~const),
$E_c$ increases with decreasing $\tau$, {\em i.e.}, with increasing
disorder. Our data show that at $\nu=1$, the effect of temperature is
qualitatively similar to the effect of disorder: with decreasing
temperature, the energy of the lowest band of extended states
decreases. At high $T$, there are no extended states below the Fermi
energy, and conductivity is due to temperature activation to the
nearest extended state. At lower temperatures, the band of extended
states crosses the Fermi level, and we observe a metallic state with
characteristic metallic temperature dependencies for both diagonal
and Hall conductivities. At still lower temperatures, the band of
extended states sinks below the Fermi energy, $\sigma_{xy}$
approaches $e^2/h$, and the system enters the QH regime.

Summarizing, we have studied the temperature dependent behavior of
the very dilute 2D electron system in silicon at fixed filling factor
$\nu=1$. We have obtained experimental evidence that at $\nu=1$, the
energy of the lowest band of extended states decreases relative to
the Fermi energy as the temperature is decreased. As a result, this
band passes through the Fermi energy, causing transitions from
insulator-like to metal-like and metal to QH-like temperature
dependencies of transport coefficients. The first of these
transitions reflects {\em delocalization} with decreasing
temperature; this is in sharp contrast to ``normal'' situation in
which lowering the temperature makes a 2DES more localized.

We acknowledge useful discussions with E.~I.~Rashba, L.~Zhang,
B.~A.~Mason, X.~C.~Xie, and D.~Shahar. This work was supported by
grants from the National Science Foundation, DMR 89-22222 and
Oklahoma EPSCoR via the LEPM, and the SERC (UK).

\begin{figure}
\caption{(a)~Diagonal and Hall resistivity of sample Si-14 in units
of $h/e^2$ vs magnetic field at three temperatures and
$n_s=9.1\times10^{10}$ cm$^{-2}$. The inset schematically shows the
expected behavior of extended states in a magnetic field
[7,8]; (b)~Inverse $\sigma_{xx}$ of sample Si-22 obtained by impedance
measurements vs magnetic field at two temperatures and
$n_s=7.8\times10^{10}$ cm$^{-2}$. The inset shows a ``conventional''
$\sigma_{xx}(B)$ for higher electron density, $n_s=1.56\times10^{11}$
cm$^{-2}$.}
\label{1}
\end{figure}
\begin{figure}
\caption{Temperature dependence of diagonal resistivity of Si-14 at
$\nu=1$ for four electron densities. Inset shows activated
temperature dependence of $\rho_{xx}$ at higher temperatures for the
same sample.}
\label{2}
\end{figure}
\begin{figure}
\caption{Temperature dependence of diagonal and Hall conductivities
for sample Si-14 at $\nu=1$ for four different $n_s$. Open symbols
correspond to $\sigma_{xy}$, closed - to $\sigma_{xx}$. Vertical
dotted lines approximately separate different kinds of temperature
behavior.}
\label{3}
\end{figure}

\begin{references}
\bibitem{floating} Experiments which are consistent with this
behavior were reported recently in Ref.\ \cite{shashkin93}.
\bibitem{hall} Hall resistivity in the presence of the reentrant
insulating phase in 2D system in silicon has been studied extensively
in Refs.\ \cite{kravchenko91,pudalov93} so we will not repeat the
details here.
\end{references}
\end{document}